\documentclass[%
 reprint,
preprintnumbers,
nofootinbib,
nobibnotes,
 amsmath,
 amssymb,
floatfix,
]{revtex4-1}
\usepackage{CJK}
\usepackage[utf8]{inputenc}
\usepackage{amsthm}
\usepackage{graphicx}
\usepackage{dcolumn}
\usepackage{bm}
\usepackage[colorlinks=true,pdfstartview=FitV,breaklinks=true]{hyperref}
\usepackage[dvipsnames,table]{xcolor}
\hypersetup{urlcolor=black, citecolor=black, linkcolor=black}

\usepackage{graphicx}
\usepackage{dcolumn}
\usepackage{bm}

\begin{document}

\preprint{CPPC-2021-07}

\title{Non-Schwarzschild Primordial Black Holes as Dark Matter in Quadratic Gravity}
\author{Yunho Kim}
 \email{yunho.kim@sydney.edu.au}
\author{Archil Kobakhidze}
\email{archil.kobakhidze@sydney.edu.au}
\affiliation{
Sydney Consortium for Particle Physics and Cosmology, \\
 School of Physics, The University of Sydney, NSW 2006, Australia 
}

\begin{abstract}
One-loop renormalised quantum effective action for gravity contains quadratic in curvature terms. We have found an approximate analytic black hole solution in quadratic gravity by keeping only the radial spherically symmetric fluctuations and dimensionally reducing the 4-dimensional (4D) theory down to the 2-dimensional (2D) dilaton gravity with a potential. The solution reduces to the Schwarzschild black hole in the limit of Einstein’s gravity, but otherwise admits non-negative Arnowitt-Deser-Misner (ADM) and positive quasi-local Misner-Sharp masses that can differ significantly. We then study the thermodynamics of such quantum corrected black holes and compute their lifetime under Hawking evaporation. We note that for some range of parameters, black holes increase in mass while emitting Hawking radiation. This pathological behaviour is related to the negative energy states that are present in quadratic gravity. We also find that the micro-lensing of non-Schwarzschild black holes could significantly deviate from the micro-lensing of their Schwarzschild counterparts. These findings have important ramifications for the phenomenology of primordial black holes (PBHs) as dark matter. In particular, the quoted constraints on PBH dark matter from micro-lensing data can be completely evaded, thus making PBHs in the mass range $\sim 10^{-12} - 10~M_{\odot}$ viable dark matter candidates.
\end{abstract}

\maketitle


\section{Introduction} \label{sec1}
Primordial black holes (PBHs) \cite{pbh, Hawking:1971ei, Carr:1974nx} are attractive candidates for the elusive dark matter. Arguably they are the least speculative candidates since they do not require extension of the known particle physics, except that their production in the early universe requires extra theoretical input. In a nutshell, a large enough fluctuation in a homogeneous radiation-dominated universe collapses into a horizon size black hole. Hence, the lighter PBH is, the earlier it has been produced.

The PBH dark matter mass is bounded from below from the stability consideration under Hawking evaporation \cite{Hawking:1974sw}. The Schwarzschild PBH of mass $m_{PBH} \gtrsim 10^{-19}~M_{\odot}$ has a lifetime longer than the age of the universe and hence can potentially serve as dark matter. However, PBH abundance is constrained by various astrophysical and cosmological observations (for recent comprehensive reviews see Refs \cite{Carr:2020gox} and \cite{Green:2020jor}). In particular, the PBHs in the mass range $\sim 10^{-12} - 10~M_{\odot}$ are severely constrained by micro-lensing experiments \cite{Allsman:2000kg, Tisserand:2006zx, Niikura:2017zjd}.

Most of the discussions on PBH dark matter phenomenology are centered around PBH properties as defined by the static Schwarzschild black hole. It has been shown recently in Refs. \cite{Boehm:2020jwd, Picker:2021jxl} that cosmological black holes (for a review see \cite{Faraoni:2018xwo}) may exhibit significantly different behaviours. Furthermore, while the Schwarzschild black hole is a vacuum solution of the Einstein classical theory of General Relativity (GR), one can expect that the solution gets modified in quantum theory. Although such quantum corrections cannot be computed to the full extent using the standard quantum field theoretic perturbation techniques, a number of explicit approximate calculations are available in the literature. For example, in Ref. \cite{Abedi:2015yga} corrections to the Schwarzschild solution due to the quantum trace anomaly have been computed. In \cite{Kazakov:1993ha}, a partial quantisation of gravity has been suggested for the background Schwarzschild geometry. The authors consider quantisation of only radial fluctuations while keeping other degrees of freedom classical and computed corrections to the Schwarzschild solution within the effective 2D dilaton gravity.

It has been well-known for a long time that the one-loop quantum contribution from matter fields results in the additional quadratic in curvature terms in the renormalised effective action for gravity \cite{Utiyama:1962sn}. The Schwarzschild metric is a solution in quantum corrected quadratic gravity. In addition, spherically symmetric and asymptotically flat non-Schwarzschild black hole solutions with positive and negative masses have been proven to exist \cite{Lu:2015cqa, Lu:2015psa}. The numerical solution for a new electrically charged black hole has been presented in \cite{Wu:2019uvq}. In Ref. \cite{Kim:2020bhg}, a new analytical charged black hole solution has been found in the theory of quadratic gravity with additional topological interactions \cite{Kobakhidze:2008em}.

In this paper, we obtain an approximate correction to the Schwarzschild black hole in quadratic gravity by following the formalism of dimensional reduction of Ref. \cite{Kazakov:1993ha}. The form of the correction is similar to the one found in \cite{Kazakov:1993ha}, except that it depends on the additional parameters related to the higher curvature terms. The resulting solution has a non-vanishing Ricci tensor, but a trivial Ricci scalar and hence falls in the category of non-Schwarzschild black hole solutions of Refs. \cite{Lu:2015cqa, Lu:2015psa}.

This quantum corrected black hole solution has significant ramifications for PBH dark matter phenomenology. We argue that the ``active" PBH mass that contributes to the local dark matter density given by the quasi-local Misner-Sharp (MS) mass \cite{Misner:1964je}, which may substantially differ from the Arnowitt-Deser-Misner (ADM) mass \cite{Arnowitt:1959ah}. Given the constraints on parameters of quadratic gravity \cite{Kim:2019sqk}, we found that PBHs with even vanishing ADM mass can be stable. Furthermore, we argue that micro-lensing is less prominent for these black holes. In fact, the micro-lensing constraints \cite{Allsman:2000kg, Tisserand:2006zx, Niikura:2017zjd} can be completely evaded and hence PBHs in the wide mass range $\sim 10^{-12} - 10~M_{\odot}$ can actually constitute 100\% of dark matter.

The rest of the paper is organised as follows. In the next section we present dimensional reduction of 4D quadratic gravity by considering spherically symmetric radial fluctuations around the Schwarzschild metric. We then obtain an approximate correction to the Schwarzschild metric by analysing the resulting dimensionally reduced theory of 2D dilaton gravity with a potential. In Sec. \ref{sec2} we study thermodynamics of quantum corrected black holes and compute their life-time. In Sec \ref{sec3} we consider micro-lensing of quantum corrected PBHs and argue that current constraints on PBH dark matter can be avoided. The last section is reserved for our conclusions. Some technicalities are presented in Appendices \ref{appenda} and \ref{appendb}.

\section{Static spherically symmetric solutions in quadratic gravity} \label{sec2}
The one-loop renormalised effective action for gravity contains terms quadratic in curvature \cite{Utiyama:1962sn}:
    \begin{equation}
        S = \int d^4x \sqrt{-g} \left[\frac{R}{16 \pi G} + \beta R^2 + \gamma R_{\mu\nu} R^{\mu\nu} \right]~,
        \label{action}
    \end{equation}
where $\sqrt{1/G}=M_{P}\simeq 1.2\cdot 10^{19}~\mathrm{GeV}$, and $\beta$ and $\gamma$ are dimensionless parameters. They cannot be calculated within the effective theory framework and must be extracted from observations. In \cite{Kim:2019sqk} we obtained the following bounds on these parameters from the LIGO/Virgo gravitational wave data with the requirement of the absence of tachyonic instabilities:
\begin{align}
    0 \leq & \enspace \> \gamma \enspace \, \lesssim 5.9\cdot 10^{76} \nonumber  \\
    -\frac{\gamma}{4} \leq & \enspace \> \beta \enspace \, \lesssim 9.8\cdot 10^{75} - \frac{\gamma}{4}~.
    \label{bounds}
\end{align}
The Schwarzschild metric (with $R=R_{\mu\nu}=0$) remains the solution for quadratic gravity, but will certainly fail if higher-order terms involving the Riemann tensor are added to (\ref{action}). We are interested here in spherically symmetric and static non-Schwarzschild solutions, which presumably are also approximate solutions for such extended effective actions. To find such approximate solutions we follow Ref. \cite{Kazakov:1993ha} and consider a general spherically symmetric metric:
\begin{equation} \label{metric}
    ds^2 = e^{\sigma(z^{+}, z^{-})} dz^{+} dz^{-} - r^2(z^{+}, z^{-}) (d\theta^2 + \sin^2\theta d\phi^2)~.
\end{equation}
Here the four-dimensional spacetime is covered by the coordinates $(z^{+}, z^{-}, \theta, \phi)$. The conformal coordinates $(z^{+}, z^{-})$ cover a 2D spacetime section, and angular coordinates $(\theta, \phi)$ cover a 2-sphere with a radius $r(z^{+}, z^{-})$ that is a function on the 2D spacetime. Note that for the Schwarzschild metric these coordinates are known as the Kruskal coordinates \cite{Kruskal:1959vx} and define the maximal extension of the Schwarzschild spacetime.

We evaluate the 4D action (\ref{action}) using the metric (\ref{metric}) and upon integrating over the 2-sphere obtain the following 2D effective action (some details of these calculations and notations are given in Appendix \ref{appenda}):
\begin{align}
    S = \int d^2z \sqrt{-\Tilde{g}} &\left[\frac{1}{8 G} \left(r^2 \Tilde{R} - 2 (\nabla r)^2 + 2 \right) \right. \nonumber \\
    &+ 2 \pi \beta \left(4 \Tilde{R} - \frac{8}{r^2} (\nabla r)^2 + \frac{8}{r^2} \Box r^2 + \frac{4}{r^2} \right) \nonumber \\
    &\left. + 2 \pi \gamma \left(\frac{2}{r^2} \Box r^2 + \frac{2}{r^2} \right) \right]~.
    \label{action2}
\end{align}
Here $\tilde g_{ab} =e^{\sigma(z^{+}, z^{-})}\left(\delta_{a+}\delta_{b-}+\delta_{a-}\delta_{b+}\right)~[a,b=+,-]$ is the 2D metric; $\tilde{R}=4e^{-\sigma} \partial_{+}\partial_{-}\sigma$ is the 2D Ricci scalar and the covariant derivatives are constructed out of $\tilde g_{ab}$. In (\ref{action2}) we have omitted inconsequential total derivative terms as well as higher-derivative terms involving the radial dilaton field $r(z^+,z^-)$. The latter approximation is justified as soon as we are interested only in solutions at distance scales bigger than the microscopic Planck scale (see Appendix \ref{appenda}). This is always true for macroscopic black holes.

Next, taking the variation of the effective action (\ref{action2}) with respect to the 2D metric $\tilde g_{ab}$ we obtain the equations of motion:
\begin{align}
    &\left(2r + \frac{64\pi G}{r} (4\beta + \gamma) \right) \nabla_{a} \nabla_{b} r + \frac{64\pi G}{r^2} (2\beta + \gamma) \nabla_{a} r \nabla_{b} r \nonumber \\
     &=\tilde  g_{ab} \left[1 + \frac{16\pi G}{r^2} (2\beta + \gamma) + \left(1 + \frac{32\pi G}{r^2} (2\beta + \gamma) \right)(\nabla r)^2 \right. \nonumber \\ 
    &\left. + \left(2r + \frac{32\pi G}{r^2} (4\beta + \gamma) \right)\Box r \right]~.
    \label{eom}
\end{align}
Using the above equation, one can demonstrate \cite{Banks:1990mk} that the 2D spacetime admits the following Killing vector:
\begin{equation}
    \xi_{a} = \varepsilon_{a b} \tilde{g}^{bc} \partial_{c} r \frac{dD(r)}{dr} e^{\int^{r} \frac{dx}{D(x)}}~,
\end{equation}  
where $\nabla_{\mu} \nabla_{\nu} D(r) = \frac{\left(2r^3 + 64\pi G (4\beta + \gamma)r \right) \nabla_{\mu} \nabla_{\nu} r}{64\pi G (2\beta + \gamma)}$. This Killing vector is orthogonal to the gradient of the dilaton field $r$, so surfaces of constant $r$ are flow lines of the Killing vector field. Thus we can take $r$ to be a spacelike coordinate and rewrite the 2D metric in the Schwarzschild form:
\begin{equation}
    ds^2 = g(r) dt^2 - \frac{1}{g(r)} dr^2~,
\end{equation}
where $(\nabla r)^2 = -g(r)$ and $\Box r = -\frac{dg(r)}{dr}$.

Tracing equation (\ref{eom}) we obtain the differential equation for the function $g(r)$:
\begin{equation}
    r \frac{dg(r)}{dr} + g(r) - \left[1 + \frac{\delta G}{r^2} \right] = 0~,
\end{equation}
where $\delta \equiv 16\pi (2\beta +\gamma)$. The solution of this equation can be readily obtained:
\begin{equation}
    g(r) = 1 - \frac{2GM}{r} - \frac{\delta G}{r^2}~,
    \label{sol}
\end{equation}
where $M$ is an integration constant. Using  (\ref{sol}), we can construct new spherically symmetric non-Schwarzschild black hole solution in 4D.

Our solution (\ref{sol}) resembles the Reissner–Nordström metric for a charged black hole. There is an important difference, however. The third term in (\ref{sol}) comes with an overall negative sign since $\delta \geq 0$. Hence, an extreme non-Schwarzschild limit $M\to 0$ is permissible, since no naked central singularity is displayed. The solution expectedly approaches the Schwarzschild metric in the limit $\delta \to 0$.

The Ricci scalar for our black hole is trivial, $R=0$, however, the Ricci tensor is not, $R_{\mu\nu}\neq 0$. Hence, this solution falls into the class of Non-Schwarzschild black holes discussed in \cite{Lu:2015cqa, Lu:2015psa}. We also note that the form of the solution (\ref{sol}) is similar to the one obtained in general relativity by quantising the corresponding effective 2D dilaton gravity \cite{Kazakov:1993ha}. In our case, the correction to the Schwarzschild black hole explicitly depends on the parameters of quadratic gravity through $\delta$ in Eq. (\ref{sol}).

\section{Mass and thermodynamics of non-Schwarzschild black holes} \label{sec3}
For non-vacuum, spherically symmetric black holes, such as the one we obtained in the previous section, it is convenient to introduce a quasi-local Misner-Sharp mass, $M_{MS}$ \cite{Misner:1964je}. The Misner-Sharp mass is a coordinate-independent quantity that physically describes the total energy within a spherical radius $r$ around the black hole. Our black hole metric then can be written as follows:
\begin{eqnarray}
    ds^2 &=& \left(1-\frac{2GM_{MS}}{r}\right)dt^2 - \left(1-\frac{2GM_{MS}}{r}\right)^{-1}dr^2 \nonumber \\
    &-&r^2\left(d\theta^2 +\sin^2\theta d\phi^2\right)~,
    \label{bhmetric}
\end{eqnarray}
where
\begin{equation}
    M_{MS}(r)=M+\frac{\delta}{2r}~.
    \label{ms1}
\end{equation}
The event horizon of the black hole (\ref{bhmetric}) is given by:
\begin{equation}
    r_h = GM\left(1+\sqrt{1+\frac{\delta}{GM^2}}\right)~.
    \label{bhhor}
\end{equation}
Then we define the black hole mass as the Misner-Sharp mass within the event horizon:
\begin{equation}
    M_{bh}=M_{MS}(r_h)=M+\frac{\delta}{2r_h}~.
    \label{ms}
\end{equation}
The above mass must be positive definite, hence $M\geq 0$. In the Schwarzschild limit $\delta\to 0$, the black hole mass becomes the ADM mass $M$, that is the mass ``observed" by an inertial observer at infinity from the black hole. In the opposite extreme non-Schwarzschild we take the ADM mass $M$ to zero and obtain:
\begin{equation}
    M_{bh}=\sqrt{\delta} M_P/2~. 
    \label{bhmass}
\end{equation}
Hence, even with a vanishing ADM mass, this black hole carries a non-vanishing quasi-local mass that defines its local gravitational properties. Baring in mind the constraints on parameters of quadratic gravity (\ref{bounds}), we find $\delta \lesssim 10^{78}$, and hence the mass of the extreme non-Schwarzschild black holes is bounded from above, $M_{bh}\lesssim 10~M_{\odot}$. If such black holes exist they are likely to have a primordial origin.

Now let us turn to the thermodynamics and stability of the above black holes under Hawking evaporation. The surface gravity and hence the Hawking temperature can be conveniently expressed through the Misner-Sharp mass (\ref{ms}) (see, e.g., discussion in \cite{Nielsen:2008kd}):
\begin{eqnarray}
    T_{H}&=&\left.\frac{1}{4\pi r}\left(1-2G\frac{dM_{MS}}{dr}\right)\right.\bigg|_{r=r_h} \nonumber \\
    &=&\frac{1}{8\pi G M_{bh}}\left(1+\frac{\delta}{4GM_{bh}^2}\right)~.
    \label{temp}
\end{eqnarray} 
Since $\delta \geq 0$, we observe that non-Schwarzschild black holes are hotter than their Schwarzschild counterparts. In the limit $M\to 0$, the Hawking temperature of the extreme non-Schwarzschild black hole is twice larger than the temperature of the same mass Schwarzschild black hole:
\begin{eqnarray}
    T_{H}&=&\frac{1}{4\pi G M_{bh}}~.
    \label{temp1}
\end{eqnarray} 
where the black hole mass is given by Eq. (\ref{bhmass}).

The expression for the black hole entropy in quadratic gravity also deviates from the canonical $A/4G$, where $A=4\pi r_h^2$ is the horizon area. Adopting calculations from Ref. \cite{Fan:2014ala}, we obtain:
\begin{align}
    \mathcal{S} &= \frac{A}{4G} \left( 1 -\frac{\beta}{GM_{bh}^2}+\frac{\gamma\delta}{8G^2M_{bh}^4}\right) \nonumber \\
    & = 4\pi G\left(M_{bh}^2+\frac{\gamma\delta}{8G^2M_{bh}^2}-\frac{\beta}{G}\right)~.
    \label{entropy}
\end{align}
The rate of the entropy change during Hawking evaporation then reads:
\begin{align}
    \frac{d\mathcal{S}}{dt}=8\pi G\left(M_{bh}-\frac{\gamma\delta}{8G^2M_{bh}^3}\right)\frac{dM_{bh}}{dt}~.   
    \label{entropyrate}
\end{align}
Next, using the thermodynamic relation $dU=T_Hd\mathcal{S}$ and the Stefan-Boltzmann law for the radiated black hole energy, $\frac{dU}{dt}=-\frac{\pi^2}{60}AT_H^4$, we derive the equation for the rate of the black hole mass change:
\begin{eqnarray}
    \frac{dM_{bh}}{dt}=-\frac{\pi^2}{30}GM_{bh}\left(1-\frac{\gamma\delta}{8G^2M_{bh}^4}\right)^{-1}T_{H}^3~.
    \label{massrate}
\end{eqnarray}
The right-hand side of the above equation becomes positive once $\frac{\gamma\delta}{8G^2M_{bh}^4}<1$, which means that the black hole mass starts increasing whilst it evaporates. This puzzling behaviour is directly related to the inherent pathology of the effective quadratic gravity. Indeed, the parameter $\gamma$ defines the mass of the spin-2 negative energy states, $m_{\pi}=M_P/\sqrt{2\gamma }$ \cite{Kim:2019sqk}. When the Hawking temperature exceeds $m_{\pi}$\footnote{Note, the temperature of a radiation comprised of negative energy quanta is formally negative.}, the massive spin-2 particles are radiated away carrying out the negative energy and thus resulting in the increase of the black hole mass. This is the regime where the quadratic truncation of the effective action (\ref{action}) is no longer valid. The resolution of the problem of negative energy states most likely requires going beyond the quadratic truncation of the effective action. Here we simply disregard this problem by setting $\gamma=0$.

We can now integrate Eq. (\ref{massrate}) with the initial black hole mass $m_{PBH}$. In the extreme non-Schwarzschild limit we obtain the following black hole lifetime:
\begin{equation}
    \tau_{PBH}=640\pi G^2m_{PBH}^3~.
    \label{life}
\end{equation}
Hence, the lifetime of the extreme non-Schwarzschild black holes is 8 times shorter than the lifetime of Schwarzschild black holes of the same mass. Correspondingly, the stability bound on the non-Schwarzschild black hole mass is twice the corresponding bound on the mass of the Schwarzschild black hole.
    
\section{Micro-lensing of non-Schwarzschild black holes} \label{sec4}
In the previous section, we have established the stability of non-Schwarzschild black holes. In this section, we turn our attention to other potential constraints on the abundance of such black holes. In particular, micro-lensing data are known to severely constrain the abundance of PBHs in the mass range $\sim 10^{-12} - 10~M_{\odot}$. To study micro-lensing for non-Schwarzschild PBHs we first compute the deflection angle of a light ray of a black hole. Following the standard steps (see, e.g. \cite{Weinberg:1972kfs} and Appendix \ref{appendb}) we obtain in the leading order of $GM_{MS}(r)/r$ the following expression for the angle:
\begin{equation}
    \Delta \varphi = \frac{4GM}{r_0}+\frac{3\pi G\delta}{4r_0^2}~,
    \label{angle}
\end{equation}
where $r_0$ is the impact parameter. The first term in Eq. (\ref{angle}) is the familiar Einstein's expression for the deflection angle. In the extreme non-Schwarzschild limit $M\to 0$, however, the deflection angle is proportional to the inverse square of the impact parameter and thus the lensing will be less prominent for such black holes. We will focus on this limit in what follows.

Now assuming the standard geometry for micro-lensing, we compute the angular position $\theta$ of the image of a source star in the presence of a black hole lens. In a special case where observer, lens and source are aligned it is related to the deflection angle (\ref{angle}) and distances $D_L$ and $D_S$ from the observer to the lens and source, respectively through the ``lens equation":
\begin{equation}
    \theta = \frac{D_S-D_L}{D_S}\Delta \varphi (\theta)~.
    \label{lens}
\end{equation}    
Note, the deflection angle itself depends on $\theta$ through the impact parameter $r_0=D_L\theta$. In the limit $M\to 0$ we obtain from Eq. (\ref{lens}):
\begin{eqnarray}
    \theta &=& \left( \frac{3\pi G^2m_{PBH}^2}{d^2}\right)^{1/3} \nonumber \\
    &=&\frac{(3\pi)^{1/3}}{2}\left(\frac{Gm_{PBH}}{d}\right)^{1/6}\theta_E~.
    \label{image}
\end{eqnarray} 
Here, we assumed $(D_S-D_L)\sim D_S\sim D_L \sim d$ and $\theta_E^2=\frac{4Gm_{PBH}}{d}$ is the canonical angle within general relativity that defines the Einstein ring. The modified Einstein ring for non-Schwarzschild black holes are several orders of magnitude smaller than $\theta_E$:
\begin{eqnarray}
    \frac{\theta}{\theta_E} &\approx& 1.4\cdot 10^{-3}~\left(\frac{m_{PBH}}{10~M_{\odot}}\right)^{1/6}\left(\frac{100~\mathrm{kPc}}{d}\right)^{1/6}~.
    \label{ratio}
\end{eqnarray} 

The existing micro-lensing experiments are not capable to detect such small black hole lenses. To quantify this we compute the optical depth for non-Schwarzschild black holes and compare it with the optical depth for Schwarzschild black holes of the same mass and abundance. The optical depth describes a probability that an image of a given star, during a specific time of observation, has been magnified significantly (magnification $>$ 1.34) due to micro-lensing. Assuming the homogeneous density of black hole lenses $n_{PBH}$, we find:
\begin{eqnarray}
    \tau &=& \pi n_{PBH}\int_0^d \theta^2(r)r^2dr \nonumber \\
    &=& \frac{3^{5/3} \pi^{2/3}}{10}\left(\frac{Gm_{PBH}}{d}\right)^{1/3}\tau_E~,
\end{eqnarray} 
where $\tau_E=2\pi Gm_{PBH} n_{PBH} d^2$ is the optical depth for the Schwarzschild black hole. Quantitatively we have:
\begin{equation}
    \frac{\tau}{\tau_E}=2.3\cdot 10^{-6}~\left(\frac{m_{PBH}}{10~M_{\odot}}\right)^{1/3}\left(\frac{100~\mathrm{kPc}}{d}\right)^{1/3}~.
\end{equation}
Hence the non-Schwarzschild black holes in the mass range $\lesssim 10~M_{\odot}$ is more than 6 orders of magnitude less likely to be detected through micro-lensing than their Schwarzschild counterparts. The available micro-lensing observations \cite{Allsman:2000kg, Tisserand:2006zx, Niikura:2017zjd} constrain the abundance of PBH to be up to $f_{PBH}\sim 10^{-3}$ fraction of dark matter abundance in the PBH mass range $\sim 10^{-12} - 10~M_{\odot}$. These experiments are not sensitive enough to constrain the abundance of non-Schwarzschild PBHs, and, hence, the non-Schwarzschild PBHs with masses $\sim 10^{-12} - 10~M_{\odot}$ are viable candidates to describe the entire dark matter in the Universe.

\section{Conclusion} \label{sec5}
In this paper, we advocated considering non- Schwarzschild PBHs as potential dark matter candidates. Such black holes may carry, distinctive from the Schwarzschild black holes, features that allow evading the existing strict bounds on PBH abundance. In particular, we have found an approximate spherically symmetric black hole solution (\ref{bhmetric}) in quadratic gravity that is different from the Schwarzschild black hole. It admits a non-zero quasi-local Misner-Sharp mass (\ref{ms}) even in the limit of vanishing ADM mass  ($M\to 0$).

The properties of our non-Schwarzschild black holes are defined by their quasi-local Misner-Sharp masses, including the Hawking temperature (\ref{temp}) and thus the rate of their evaporation (\ref{massrate}). We have found that the lifetime of extreme non-Schwarzschild black holes ($M\to 0$ in Eq. (\ref{ms})) is 8 times shorter than those of Schwarzschild black holes of the same mass. Correspondingly, the stability bound on the mass for non-Schwarzschild black holes moves up only by a factor of two relative to the standard bound quoted for Schwarzschild black holes. Furthermore, we have demonstrated that the extreme non-Schwarzschild black holes deflect the passing light at much smaller angles  (\ref{ratio}) than their Schwarzschild counterparts. This allows us to completely evade the micro-lensing constraints, rendering non-Schwarzschild PBHs in a wide mass range $\sim 10^{-12} - 10~M_{\odot}$ as viable dark matter candidates.

\begin{acknowledgments}
The work of AK was partially supported by the Australian Research Council (grant DP210101636) and the Shota Rustaveli National Science Foundation of Georgia (SRNSFG) through the grant DI-18-335.
\end{acknowledgments}

\appendix
\section{Quadratic in curvature terms}  \label{appenda}
For the readers' convenience, below we give details of the calculations presented in the main text.

The non-zero components of the Ricci tensor for the metric (\ref{metric}) are:
    \begin{align}
       & R_{+-} = \partial_{+} \partial_{-} \sigma + \partial_{+} \partial_{-} \ln{r^2} + \frac{1}{2} \partial_{+} \ln{r^2} \partial_{-} \ln{r^2}~, \nonumber \\
       &R_{\pm\pm} = \partial^2_{\pm} \ln{r^2} + \partial_{\pm} \ln{r^2} \partial_{\pm} \ln{r^2} - \partial_{\pm} \sigma \partial_{\pm} \ln{r^2}~, 
         \nonumber \\
       & R_{\theta\theta} = -1 - 2 e^{-\sigma} \partial_{+} \partial_{-} r^2~,~~
        R_{\phi\phi} = \sin^2 \theta R_{\theta\theta}~.
    \end{align}
The Ricci scalar, the Ricci scalar squared and the Ricci curvature squared can then be computed as follows:
    \begin{align}
       & R = \tilde{R} - \frac{2}{r^2} (\nabla r)^2 + \frac{2}{r^2} \Box r^2 + \frac{2}{r^2}~, \nonumber \\
       & R^2  = \frac{4}{r^2} \tilde{R} - \frac{8}{r^4} (\nabla r)^2 + \frac{8}{r^4} \Box r^2 + \frac{4}{r^4} +\mathcal{O}\left((\nabla r)^4\right)~,\nonumber \\
       & R_{\mu\nu} R^{\mu\nu} = \frac{2}{r^4} \Box r^2 + \frac{2}{r^4}+\mathcal{O}\left((\nabla r)^4\right)~. 
       \label{a1}
    \end{align}
In the last two equations, we have omitted terms with higher than two derivatives of $r$. These are terms like  $(\nabla r)^4/r^4$, $(\Box r^2)^2/r^4$, $(\nabla r)^2 \Box r^2/r^4$. The following estimate justifies this omission. We compare e.g. the omitted term $(\nabla r)^4/r^4$ with the leading order term with two derivatives  $(\nabla r)^2/G r^2$. Since,  $(\nabla r)^2\propto g(r)=1-2GM_{MS}(r)/r$, we conclude that the higher-derivative terms are always sub-dominant whenever $r>1/M_{P}$ holds. 

Substituting (\ref{a1}) into (\ref{action}) and performing angular integration, one arrives at the 2D effective action in Eq. (\ref{action2}). 

\section{Evaluation of the deflection angle} \label{appendb}
A light ray deflects from a straight line trajectory in the presence of a black hole. The deflection angle is given by the formulae \cite{Weinberg:1972kfs}:
\begin{equation}
\Delta \varphi = 2\vert \varphi(r_0)-\varphi_{\infty}\vert -\pi~,
\end{equation} 
where $r_0$ is the distance of the closest approach and $\varphi_{\infty}$ is an angle at which incident light approaches the black hole. Integrating the relevant geodesic equation one obtains:
\begin{equation}
\Delta \varphi = 2\int_{r_0}^{\infty} \frac{dr}{r\sqrt{g(r)}}\left[\left(\frac{r}{r_0}\right)^2\frac{g(r_0)}{g(r)}-1\right]^{-1/2} -\pi~,
\end{equation} 
where $g(r)=1-\frac{2GM_{MS}(r)}{r}$ (see Eqs. (\ref{bhmetric}) and (\ref{ms1})). The above integral is evaluated in the first-order of $\frac{GM_{MS}(r)}{r}$:
\begin{eqnarray}
\Delta \varphi &\simeq& 2\int_{r_0}^{\infty}\frac{dr}{r\sqrt{\left(\frac{r}{r_0}\right)^2-1}} \left[1+\frac{GM}{r}+\frac{GM}{r_0(1-r_0/r)} \right. \nonumber \\
&+& \left.\frac{G\delta}{2}\left(\frac{1}{r^2}+\frac{1}{r_0^2}\right)\right] - \pi=\frac{4GM}{r_0}+\frac{3\pi G\delta}{4r_0^2}~.
\end{eqnarray}
This is the result displayed in Eq. (\ref{angle}).


\begin{thebibliography}{999}
\bibitem{pbh} 
  Ya. B. Zel’dovich and I. D. Novikov, ``The Hypothesis of Cores Retarded During Expansion and the Hot Cosmological Model,'' Sov. Astron. {\bf 10}, 602 (1967).
  
\bibitem{Hawking:1971ei} 
  S.~Hawking,
  ``Gravitationally collapsed objects of very low mass,''
  Mon.\ Not.\ Roy.\ Astron.\ Soc.\  {\bf 152}, 75 (1971).
  
\bibitem{Carr:1974nx} 
  B.~J.~Carr and S.~W.~Hawking,
  ``Black holes in the early Universe,''
  Mon.\ Not.\ Roy.\ Astron.\ Soc.\  {\bf 168}, 399 (1974).
  
\bibitem{Hawking:1974sw}
S.~W.~Hawking,
``Particle Creation by Black Holes,''
Commun. Math. Phys. \textbf{43}, 199-220 (1975)
[erratum: Commun. Math. Phys. \textbf{46}, 206 (1976)]
  
\bibitem{Carr:2020gox} 
  B.~Carr, K.~Kohri, Y.~Sendouda and J.~Yokoyama,
  ``Constraints on Primordial Black Holes,''
  arXiv:2002.12778 [astro-ph.CO].
  
\bibitem{Green:2020jor}
A.~M.~Green and B.~J.~Kavanagh,
``Primordial Black Holes as a dark matter candidate,''
[arXiv:2007.10722 [astro-ph.CO]].

\bibitem{Allsman:2000kg}
R.~A.~Allsman \textit{et al.} [Macho],
``MACHO project limits on black hole dark matter in the 1-30 solar mass range,''
Astrophys. J. Lett. \textbf{550}, L169 (2001)
[arXiv:astro-ph/0011506 [astro-ph]].

\bibitem{Tisserand:2006zx}
P.~Tisserand \textit{et al.} [EROS-2],
``Limits on the Macho Content of the Galactic Halo from the EROS-2 Survey of the Magellanic Clouds,''
Astron. Astrophys. \textbf{469}, 387-404 (2007)
[arXiv:astro-ph/0607207 [astro-ph]].

\bibitem{Niikura:2017zjd}
H.~Niikura, M.~Takada, N.~Yasuda, R.~H.~Lupton, T.~Sumi, S.~More, T.~Kurita, S.~Sugiyama, A.~More and M.~Oguri, \textit{et al.}
``Microlensing constraints on primordial black holes with Subaru/HSC Andromeda observations,''
Nature Astron. \textbf{3}, no.6, 524-534 (2019)
[arXiv:1701.02151 [astro-ph.CO]].

 
\bibitem{Boehm:2020jwd}
C.~Boehm, A.~Kobakhidze, C.~A.~J.~O'Hare, Z.~S.~C.~Picker and M.~Sakellariadou,
``Eliminating the LIGO bounds on primordial black hole dark matter,''
JCAP \textbf{03}, 078 (2021)
[arXiv:2008.10743 [astro-ph.CO]];
``Comment on: Cosmological black holes are not described by the Thakurta metric,''
arXiv:2105.14908 [astro-ph.CO].
 
\bibitem{Picker:2021jxl}
Z.~S.~C.~Picker,
[arXiv:2103.02815 [astro-ph.CO]].

\bibitem{Faraoni:2018xwo}
V.~Faraoni,
``Embedding black holes and other inhomogeneities in the universe in various theories of gravity: a short review,''
Universe \textbf{4}, no.10, 109 (2018)
[arXiv:1810.04667 [gr-qc]].

\bibitem{Abedi:2015yga}
J.~Abedi and H.~Arfaei,
``Obstruction of black hole singularity by quantum field theory effects,''
JHEP \textbf{03}, 135 (2016)
[arXiv:1506.05844 [gr-qc]].

\bibitem{Kazakov:1993ha}
D.~I.~Kazakov and S.~N.~Solodukhin,
``On Quantum deformation of the Schwarzschild solution,''
Nucl. Phys. B \textbf{429}, 153-176 (1994)
[arXiv:hep-th/9310150 [hep-th]].

\bibitem{Utiyama:1962sn}
R.~Utiyama and B.~S.~DeWitt,
J. Math. Phys. \textbf{3}, 608-618 (1962)


\bibitem{Lu:2015cqa}
H.~L\"u, A.~Perkins, C.~N.~Pope and K.~S.~Stelle,
``Black Holes in Higher-Derivative Gravity,''
Phys. Rev. Lett. \textbf{114}, no.17, 171601 (2015)
[arXiv:1502.01028 [hep-th]].

\bibitem{Lu:2015psa}
H.~L\"u, A.~Perkins, C.~N.~Pope and K.~S.~Stelle,
``Spherically Symmetric Solutions in Higher-Derivative Gravity,''
Phys. Rev. D \textbf{92}, no.12, 124019 (2015)
[arXiv:1508.00010 [hep-th]].

\bibitem{Wu:2019uvq}
C.~Wu, D.~C.~Zou and M.~Zhang,
``Charged black holes in the Einstein-Maxwell-Weyl gravity,''
Nucl. Phys. B \textbf{952}, 114942 (2020)
[arXiv:1904.10193 [gr-qc]].

\bibitem{Kim:2020bhg}
Y.~Kim and A.~Kobakhidze,
``Topologically induced black hole charge and its astrophysical manifestations,''
[arXiv:2008.04506 [gr-qc]].


\bibitem{Kobakhidze:2008em}
A.~Kobakhidze and B.~H.~J.~McKellar,
``(De)quantization of black hole charges,''
Class. Quant. Grav. \textbf{25}, 195002 (2008)
[arXiv:0803.3680 [hep-th]].

\bibitem{Misner:1964je}
C.~W.~Misner and D.~H.~Sharp,
``Relativistic equations for adiabatic, spherically symmetric gravitational collapse,''
Phys. Rev. \textbf{136}, B571-B576 (1964)

\bibitem{Arnowitt:1959ah}
R.~L.~Arnowitt, S.~Deser and C.~W.~Misner,
``Dynamical Structure and Definition of Energy in General Relativity,''
Phys. Rev. \textbf{116}, 1322-1330 (1959)

\bibitem{Kim:2019sqk}
Y.~Kim, A.~Kobakhidze and Z.~S.~C.~Picker,
``Probing Quadratic Gravity with Binary Inspirals,''
Eur. Phys. J. C \textbf{81}, no.4, 362 (2021)
[arXiv:1906.12034 [gr-qc]].

\bibitem{Kruskal:1959vx}
M.~D.~Kruskal,
``Maximal extension of Schwarzschild metric,''
Phys. Rev. \textbf{119}, 1743-1745 (1960)

\bibitem{Banks:1990mk}
T.~Banks and M.~O'Loughlin,
``Two-dimensional quantum gravity in Minkowski space,''
Nucl. Phys. B \textbf{362}, 649-664 (1991)

\bibitem{Nielsen:2008kd}
A.~B.~Nielsen and D.~H.~Yeom,
``Spherically symmetric trapping horizons, the Misner-Sharp mass and black hole evaporation,''
Int. J. Mod. Phys. A \textbf{24}, 5261-5285 (2009)
[arXiv:0804.4435 [gr-qc]].

\bibitem{Fan:2014ala}
Z.~Y.~Fan and H.~Lu,
``Thermodynamical First Laws of Black Holes in Quadratically-Extended Gravities,''
Phys. Rev. D \textbf{91}, no.6, 064009 (2015)
[arXiv:1501.00006 [hep-th]].

\bibitem{Weinberg:1972kfs}
S.~Weinberg,
``Gravitation and Cosmology: Principles and Applications of the General Theory of Relativity,''  John Wiley \& Sons (1972).
\end{thebibliography}
\end{document}